\documentclass[conference]{IEEEtran}
\IEEEoverridecommandlockouts

\usepackage{bm}
\usepackage{subfigure}
\usepackage{algorithm}

\usepackage{cite}
\usepackage{amsmath,amssymb,amsfonts}
\usepackage{algorithmic}
\usepackage{graphicx}
\usepackage{textcomp}
\usepackage{xcolor}
\def\BibTeX{{\rm B\kern-.05em{\sc i\kern-.025em b}\kern-.08em
    T\kern-.1667em\lower.7ex\hbox{E}\kern-.125emX}}
\begin{document}

\title{Online Architecture Search for Compressed Sensing \\based on Hypergradient Descent\\
\thanks{This work was supported by JSPS KAKENHI Grant-in-Aid 
for Young Scientists Grant Number JP23K13334 (to A. N.-K.) 
and JST CRONOS Japan Grant Number JPMJCS25N5 (to T. W).}
}

\author{\IEEEauthorblockN{Ayano Nakai-Kasai}
\IEEEauthorblockA{\textit{Graduate School of Engineering} \\
\textit{Nagoya Institute of Technology}\\
Nagoya, Japan \\
nakai.ayano@nitech.ac.jp}
\and
\IEEEauthorblockN{Yusuke Nakane}
\IEEEauthorblockA{\textit{Graduate School of Engineering} \\
\textit{Nagoya Institute of Technology}\\
Nagoya, Japan \\
y.nakane.547@stn.nitech.ac.jp}
\and
\IEEEauthorblockN{Tadashi Wadayama}
\IEEEauthorblockA{\textit{Graduate School of Engineering} \\
\textit{Nagoya Institute of Technology}\\
Nagoya, Japan \\
wadayama@nitech.ac.jp}
}

\maketitle

\begin{abstract}
    AS-ISTA (Architecture Searched-Iterative Shrinkage Thresholding Algorithm) and AS-FISTA (AS-Fast ISTA) are compressed sensing algorithms 
    introducing structural parameters to ISTA and FISTA 
    to enable architecture search within the iterative process.
    The structural parameters are determined using deep unfolding, 
    but this approach requires training data and the large overhead of training time.
    In this paper, we propose HGD-AS-ISTA (Hypergradient Descent-AS-ISTA) and HGD-AS-FISTA that use hypergradient descent, 
    which is an online hyperparameter optimization method, 
    to determine the structural parameters.
    Experimental results show that the proposed method improves performance of the conventional ISTA/FISTA 
    while avoiding the need for re-training when the environment changes.
\end{abstract}


\section{Introduction}
Compressed sensing has attracted attention in various fields such as medical image processing and wireless communications \cite{chanel,mimo}.
ISTA (Iterative Shrinkage Thresholding Algorithm) \cite{ista}
is a representative solution that reconstructs sparse signals 
by alternately applying a gradient descent step and a shrinkage step.
FISTA (Fast ISTA) \cite{fista} is a variant of ISTA that adds a momentum term for accelerating convergence.
Deep unfolding \cite{lista,modelbase} can effectively learn hyperparameters included in ISTA and FISTA by using training data 
and thereby improve the convergence speed and estimation accuracy.

Conventional deep unfolding approaches typically employ a fixed algorithm architecture and focus solely on learning hyperparameters.
AS-ISTA/FISTA (Architecture Searched-ISTA/FISTA) \cite{asista,asfista}, proposed recently,
achieve further performance improvement by incorporating the idea of architecture search in neural networks \cite{darts,automl}.
For each iteration in AS-ISTA/FISTA,
structural parameters are introduced that determine which step is applied.
For example, 
in the early iterations where the estimation accuracy is still low, 
it may be beneficial to apply more gradient descent steps than shrinkage steps in order to approach the solution more rapidly.
The structural parameters are learned by using deep unfolding similar to the hyperparameters.

However, the architecture search based on deep unfolding requires a large amount of training data and substantial training time.
The optimal algorithm architecture may depend on 
the statistics of the measurement matrix or system parameters, 
i.e., the execution environment. 
It is necessary to retrain the algorithm architecture for each environment 
if the algorithm is used in a time-varying environment such as wireless communications \cite{milli}.
Although it is known that the amount of training data and required training time are typically lower than those of black-box optimization based on neural networks,
they remain significant challenges in practical problems.
Therefore, it is necessary to find a method that can determine the algorithm architecture in a computationally efficient manner, 
requiring as little training data and training time as possible.

On the other hand, deep unfolding can be interpreted as an automated hyperparameter optimization method. 
Automated hyperparameter optimization methods 
have recently attracted attention in machine learning,
and a lot of efficient methods from a computational complexity viewpoint have been proposed \cite{automl}.
HGD (Hypergradient Descent) \cite{hgd} is 
one such method that optimizes the learning rate of gradient descent algorithms online.
By using the hypergradient, which is the gradient of the objective function with respect to the learning rate,
it enables updating the learning rate with low computational cost simultaneously with the execution of the gradient descent step. 
This means that the learning rate can be updated online, independent of training data.

Based on the idea that 
HGD can be applied to algorithms other than gradient descent and to hyperparameters other than the learning rate,
as long as the hypergradient is computable,
this paper proposes online architecture search methods for ISTA and FISTA.
The main contributions of this paper are listed below:
\begin{itemize}
    \item We propose HGD-AS-ISTA and HGD-AS-FISTA, 
which apply HGD to the parameter learning of AS-ISTA and AS-FISTA, respectively.
    \item The proposed methods can determine the algorithm architecture online without the need for training data and pre-training time. 
    \item Computer experiments show that the proposed methods exhibit better performance than conventional ISTA and FISTA 
    and have higher adaptability to the execution environment than AS-ISTA and AS-FISTA.
\end{itemize}

\section{Preliminaries}
\subsection{Notation}
This paper uses the following notation.
For a vector $\bm{x}\in\mathbb{R}^N$, $\|\bm{x}\|_1$ and $\|\bm{x}\|_2$ denote the $\ell_1$ and $\ell_2$ norms, respectively,
and $\mathrm{diag[\bm{x}]}$ denotes a diagonal matrix with the vector in square brackets as its diagonal elements.
A vector of all ones is denoted by $\mathbf{1}$.
For a scalar $q\in\mathbb{R}$, $\sigma(q)$ is the sigmoid function $\sigma(q) = {1}/{(1+e^{-q})}$ 
and the soft-thresholding function $S_{\tau}(q)$ is defined as $S_{\tau}(q) = \mathrm{sign}(q)\max(|q| - \tau, 0)$.
When the argument is a vector, these functions output a vector obtained by applying the function to each element.

\subsection{ISTA and FISTA}
For the compressed sensing problem setup, consider the LASSO (Least Absolute Shrinkage and Selection Operator) optimization problem \cite{cs_donoho}:
\begin{equation}
\min_{\bm{x}} \frac{1}{2}\|\bm{y} - \bm{A}\bm{x}\|_2^2 + \lambda \|\bm{x}\|_1,
\label{eq:model}
\end{equation}
where $\bm{y}=\bm{Ax}^\star+\bm{v}\in\mathbb{R}^M$ is the observation vector, $\bm{A}\in\mathbb{R}^{M\times N}$ is the measurement matrix, $\lambda > 0$ is the regularization parameter,
$\bm{x}^\star\in\mathbb{R}^N$ is the true sparse signal, $\bm{v}\in\mathbb{R}^M$ is the observation noise, and $M<N$.

ISTA (Iterative Shrinkage Thresholding Algorithm) is the iterative algorithm obtained by applying the proximal gradient method to LASSO.
With learning rate $\gamma > 0$, the ISTA update equations at the $t$-th iteration ($t=0,1,2,\ldots, T-1$) are
\begin{equation}
\bm{r}^{(t)} = \bm{x}^{(t)} - \gamma \bm{A}^\top (\bm{A} \bm{x}^{(t)} - \bm{y}),
\label{eq:gd}
\end{equation}
\begin{equation}
\bm{x}^{(t+1)} = S_{\lambda\gamma}\!\left(\bm{r}^{(t)}\right).
\label{eq:prox}
\end{equation}

FISTA is a variant of ISTA that introduces a momentum term for accelerating convergence.
FISTA adds the following update rules to the ISTA update equations, 
and the estimated value is $\bm{z}^{(t+1)}$ instead of $\bm{x}^{(t+1)}$:
\begin{equation}
  s^{(t+1)}=\frac{1+\sqrt{1+4(s^{(t)})^2}}{2},
  \label{eq:s_t}
\end{equation}
\begin{equation}
  \bm{z}^{(t+1)}=\bm{x}^{(t+1)}+\frac{s^{(t)}-1}{s^{(t+1)}}(\bm{x}^{(t+1)}-\bm{x}^{(t)}).
  \label{eq:z_t}
\end{equation}

\subsection{Deep Unfolded ISTA and FISTA}

DU-ISTA (Deep Unfolded ISTA) is
an algorithm that applies deep unfolding-based learning to the learning rate $\gamma$ contained in the ISTA \eqref{eq:gd} and \eqref{eq:prox} \cite{modelbase}.
For each iteration $t=0,\dots,T-1$, the learning rate $\gamma^{(t)}$ is introduced as a learnable parameter, 
and the update equation is modified as $\bm{x}^{(t+1)}
= S_{\gamma^{(t)}\lambda}\left(
\bm{x}^{(t)} - \gamma^{(t)} \bm{A}^{\top}\left(\bm{A}\bm{x}^{(t)} - \bm{y}\right)
\right)$.
For DU-FISTA, a similar modification is introduced into the algorithm of FISTA.
The parameters $\gamma^{(t)}$ are learned using training data consisting of pairs of observations and true sparse signals, 
and using stochastic gradient descent to minimize the MSE (mean squared error) loss function.


Incremental learning\cite{modelbase} is applied for training in this paper.
Rather than fixing the number of network layers (iterations) at the maximum,
this is a method that progressively increases the number of layers from 1 to the maximum while advancing training,
and it is expected to improve training stability and convergence.







\subsection{AS-ISTA and AS-FISTA}


AS-ISTA\cite{asista} and AS-FISTA\cite{asfista} are approaches that explore the algorithm architecture by applying operation selection based on the idea of DARTS (Differentiable Architecture Search)\cite{darts}, a neural network architecture search method, to ISTA and FISTA, respectively.
The iterations of AS-ISTA are expressed as
\begin{equation}
  \bm{r}^{(t)} = w_{r,1}^{(t)} f^{(t)}(\bm{x}^{(t)}) + w_{r,2}^{(t)} g^{(t)}(\bm{x}^{(t)}),
  \label{eq:asista_r}
\end{equation}
\begin{equation}
  \bm{x}^{(t+1)} = w_{x,1}^{(t)} f^{(t)}(\bm{r}^{(t)}) + w_{x,2}^{(t)} g^{(t)}(\bm{r}^{(t)}),
  \label{eq:asista_x}
\end{equation}
where $f^{(t)}$ and $g^{(t)}$ correspond to the gradient descent step \eqref{eq:gd} and shrinkage step \eqref{eq:prox} of ISTA, respectively, 
i.e.,
\begin{equation}
  f^{(t)}(\bm{x}) = \bm{x} - \gamma^{(t)} \bm{A}^{\top}(\bm{A}\bm{x} - \bm{y}),
  \label{eq:f_t}
\end{equation}
\begin{equation}
  g^{(t)}(\bm{x}) = S_{\gamma^{(t)}\lambda}(\bm{x}).
\end{equation}
The introduction of the weights $w_{\cdot,k}^{(t)}$ enables the selection of an appropriate step at each iteration stage, 
allowing for convergence acceleration and improved estimation accuracy 
compared to DU-ISTA/FISTA that do not include algorithm selection.

AS-FISTA adds the following update rules to the AS-ISTA update equations: \eqref{eq:s_t} and 
\begin{equation}
  \bm{z}^{(t+1)}=w_{z,1}^{(t)}h(\bm{x}^{(t+1)},\bm{x}^{(t)},s^{(t+1)},s^{(t)})+w_{z,2}^{(t)}\bm{x}^{(t+1)},
  \label{eq:asfista_z}
\end{equation}
where $h(\bm{x}^{(t+1)},\bm{x}^{(t)},s^{(t+1)},s^{(t)})=\bm{x}^{(t+1)}+((s^{(t)}-1)/s^{(t+1)})(\bm{x}^{(t+1)}-\bm{x}^{(t)})$.

The $w_{r,k}^{(t)},w_{x,k}^{(t)},w_{z,k}^{(t)} \in \mathbb{R} \ (k = 1,2)$ in the update rules are
defined using structural parameters $\beta_{r,k}^{(t)},\beta_{x,k}^{(t)},\beta_{z,k}^{(t)}$ as
\begin{equation}
w_{\cdot,k}^{(t)} = \frac{\exp(\beta_{\cdot,k}^{(t)})}{\sum_{k'=1}^{2} \exp(\beta_{\cdot,k'}^{(t)})}, \quad (\cdot \in \{r, x, z\}, k=1,2).
\label{eq:expr}
\end{equation}
This formulation encourages the weights $w_{\cdot,k}^{(t)}$ to take values $0$ or $1$, which is desirable for the architecture selection.
By definition, we always have $\sum_{k=1}^{2} w_{\cdot,k}^{(t)}=1$.
Deep unfolding-based learning is applied alternately with respect to the learning rate $\gamma^{(t)}$ and structural parameters $\beta_{\cdot,k}^{(t)}$.

\subsection{Hypergradient Descent}

HGD \cite{hgd} is a framework for updating the learning rate of gradient descent online.
Here we describe the case of simple gradient descent.
Consider the optimization parameter $\bm{\theta} \in \mathbb{R}^d$ and the objective function $f(\bm{\theta})$ with respect to the parameter.
The update rule using learning rate $\alpha$ is $\bm{\theta}_{t} = \bm{\theta}_{t-1} - \alpha \nabla f(\bm{\theta}_{t-1})$.
Let the learning rate be an iteration-dependent variable $\alpha_t$.
It is updated by 
\begin{equation}
\alpha_{t} = \alpha_{t-1} - \gamma_\alpha \frac{\partial f(\bm{\theta}_{t-1})}{\partial \alpha_{t-1}}
\label{eq:alpha_update}
\end{equation}
where $\gamma_\alpha > 0$ is the meta learning rate.

The gradient with respect to $\alpha$ can be computed using the chain rule 
and substituting the update rule for $\bm{\theta}_{t-1}=\bm{\theta}_{t-2} - \alpha_{t-1} \nabla f(\bm{\theta}_{t-2})$ as 
\begin{equation}
\frac{\partial f(\bm{\theta}_{t-1})}{\partial \alpha_{t-1}}
= \frac{\partial f(\bm{\theta}_{t-1})}{\partial \bm{\theta}_{t-1}}
 \frac{\partial \bm{\theta}_{t-1}}{\partial \alpha_{t-1}}
= -\nabla f(\bm{\theta}_{t-1})^\mathrm{T} \nabla f(\bm{\theta}_{t-2}).
\label{eq:alphat}
\end{equation}
The hypergradient can be computed as the inner product of gradients at two consecutive steps, requiring low computational cost and only additional memory to store the previous gradient.



\section{HGD-AS-ISTA and HGD-AS-FISTA}

The idea of HGD can be applied to algorithms other than gradient descent and to hyperparameters other than the learning rate.
In this paper, to eliminate the need for training data and training time in the deep unfolding used for parameter learning in AS-ISTA \cite{asista} and AS-FISTA \cite{asfista},
we propose HGD-AS-ISTA and HGD-AS-FISTA, which introduce HGD for parameter learning instead of deep unfolding.
This section describes the overview of the proposed methods 
and the specific calculations of the hypergradients are shown in Appendices.

In each iteration of HGD-AS-ISTA, 
the parameters $\alpha^{(t)}\in\{\beta_{r,1}^{(t)}, \beta_{r,2}^{(t)}, \beta_{x,1}^{(t)}, \beta_{x,2}^{(t)}, \gamma^{(t)}\}$ are updated by HGD after the forward process \eqref{eq:asista_r} and \eqref{eq:asista_x} of ISTA.
The update rules of the parameters are expressed as
\begin{equation}
\alpha^{(t+1)} = \alpha^{(t)} - \eta_\cdot \frac{\partial \tilde{J}(\bm{x}^{(t+1)})}{\partial \alpha^{(t)}},
\label{eq:hgd_asista_beta_r}
\end{equation}
where $\eta_\cdot>0 \ (\cdot\in\{r,x,\gamma\})$ are the meta learning rates.
It should be noted that the objective function $J(\bm{x}) = \,\|\bm{y} - \bm{Ax}\|_2^2/2 + \lambda \|\bm{x}\|_1$ of the optimization problem \eqref{eq:model} is not differentiable in the second term,
so we use the differentiable surrogate objective function $\tilde{J}(\bm{x}) = \,\|\bm{y} - \bm{Ax}\|_2^2/2 + \lambda\widetilde{\|\bm{x}\|_1}$.
The term $\widetilde{\|\bm{x}\|_1}$ is a differentiable approximation of the $\ell_1$ norm, which is defined as 
$\widetilde{\|\bm{x}\|_1} = \sum_{i=1}^N\left((\log(1+e^{p x_i}) + \log(1+e^{-p x_i}) - 2\log 2)/p\right)$,
where $p > 0$ controls the smoothness of the approximation.

For HGD-AS-FISTA, 
the parameters $\alpha^{(t)}\in\{\beta_{r,1}^{(t)}, \beta_{r,2}^{(t)}, \beta_{x,1}^{(t)}, \beta_{x,2}^{(t)}, \beta_{z,1}^{(t)}, \beta_{z,2}^{(t)}, \gamma^{(t)}\}$
are updated as 
\begin{equation}
\alpha^{(t+1)} = \alpha^{(t)} - \eta_\cdot \frac{\partial \tilde{J}(\bm{z}^{(t+1)})}{\partial \alpha^{(t)}},
\label{eq:hgd_asfista_beta_z}
\end{equation}
where $\eta_\cdot>0 \ (\cdot\in\{r,x,z,\gamma\})$ are the meta learning rates.

Algorithm~\ref{algo:hgd_asista} summarizes the proposed algorithms.
These update rules enable online parameter updates.
The proposed methods use STE (Straight-Through Estimator) \cite{ste} 
where the rounded weights are used in the forward process 
but the values before rounding are used for the hypergradient calculation to avoid vanishing gradients.
The obtained structural parameters are substituted for \eqref{eq:expr}, 
and then rounded to either 0 or 1, enabling algorithmic architecture selection.

The specific hypergradient calculations using the chain rule are shown in Appendices.
The only requirements for the calculation are $\bm{x}^{(t)},\bm{r}^{(t)},\bm{x}^{(t+1)},\beta_{\cdot,k}^{(t)},\gamma^{(t)}$ for HGD-AS-ISTA 
and additionally $s^{(t)},s^{(t+1)},\bm{z}^{(t+1)},\beta_{z,k}^{(t)}$ for HGD-AS-FISTA.
These are the components used at $t$-th iteration and no additional memory consumption is required.
Note that the calculation can be also performed using automatic differentiation 
but it takes additional time and consumes memory due to the construction of the computational graph.

\begin{algorithm}[tb]
  \floatname{algorithm}{Algorithm}
  \renewcommand{\algorithmicrequire}{\textbf{Input:}}
  \renewcommand{\algorithmicensure}{\textbf{Output:}}
  \caption{HGD-AS-ISTA/FISTA}
  \label{algo:hgd_asista}
  \begin{algorithmic}[1]
      \REQUIRE $\bm{x}^{(0)} = \bm{0}$, $\bm{y}$, $\bm{A}$, $\lambda$, $\alpha^{(0)}$, $\eta_\cdot$, $T$
      \FOR{$t=0, 1, \ldots, T-1$}
          \STATE Compute $w_{\cdot,k}^{(t)}$ using \eqref{eq:expr} and round them to $0$ or $1$
          \STATE $\bm{r}^{(t)} = w_{r,1}^{(t)} f^{(t)}(\bm{x}^{(t)}) + w_{r,2}^{(t)} g^{(t)}(\bm{x}^{(t)})$
          \STATE $\bm{x}^{(t+1)} = w_{x,1}^{(t)} f^{(t)}(\bm{r}^{(t)}) + w_{x,2}^{(t)} g^{(t)}(\bm{r}^{(t)})$
          \IF{HGD-AS-FISTA}
            \STATE $s^{(t+1)}=(1+\sqrt{1+4(s^{(t)})^2})/2$
            \STATE $\bm{z}^{(t+1)}\!=\!w_{z,1}^{(t)}h(\bm{x}^{(t+1)},\bm{x}^{(t)},s^{(t+1)},s^{(t)})\!+\!w_{z,2}^{(t)}\bm{x}^{(t+1)}$
          \ENDIF
          \IF{HGD-AS-ISTA for $\alpha^{(t)}\in\{\beta_{r,k}^{(t)}, \beta_{x,k}^{(t)}, \gamma^{(t)}\}$}
            \STATE $\alpha^{(t+1)} = \alpha^{(t)} - \eta_\cdot \frac{\partial \tilde{J}(\bm{x}^{(t+1)})}{\partial \alpha^{(t)}}$
          \ENDIF
          \IF{HGD-AS-FISTA for $\alpha^{(t)}\in\{\beta_{r,k}^{(t)}, \beta_{x,k}^{(t)}, \beta_{z,k}^{(t)}, \gamma^{(t)}\}$}
            \STATE $\alpha^{(t+1)} = \alpha^{(t)} - \eta_\cdot \frac{\partial \tilde{J}(\bm{z}^{(t+1)})}{\partial \alpha^{(t)}}$
          \ENDIF
      \ENDFOR
      \ENSURE $\bm{x}^{(T)}$ or $\bm{z}^{(T)}$
  \end{algorithmic}
\end{algorithm}

\section{Simulation Results}
We evaluate the performance of the proposed method on the computer with CPU Intel Xeon w7-2495X and GPU NVIDIA RTX A4000.
We assumed train-test inconsistent setting 
where the measurement matrices $\bm{A}$ whose components follow an i.i.d. standard Gaussian distribution were used for pre-training of AS-ISTA/FISTA, 
and evaluated 
where $\bm{A}$ follows the correlated Gaussian matrix with correlation coefficient $\rho=0.5$.
The parameters were set to $(M,N,\lambda,T)=(75,150,10,40)$.
The sparse original signal $\bm{x}_i^\star$ followed a Bernoulli-Gaussian distribution with mean 0 and variance 1, and the ratio of nonzero components was set to $0.08$.
The noise $\bm{v}_i$ followed an i.i.d. Gaussian distribution with mean 0 and variance $0.1$. 
For evaluation,
100 measurement matrices $\bm{A}$ were generated, and 100 sparse original signals $\bm{x}_i$ were generated for each $\bm{A}$.
As the evaluation metric, we used the following arithmetic mean for calculating MSE: 
$\sum_{i=1}^{N_{\mathrm{data}}} \left\| \bm{x}_i^{(T)} - \bm{x}_i^\star \right\|_2^2/N_{\mathrm{data}}$,
where $N_{\mathrm{data}}$ is the number of evaluation signals.

For comparison with the proposed method, we used ISTA and FISTA with a fixed learning rate, 
and AS-ISTA/FISTA for both pre-trained on i.i.d. matrices and re-trained on new environment cases.
For ISTA and FISTA, $\gamma$ was set to $\gamma=1/s_{\max}^2$,
where $s_{\max}$ is the maximum singular value of the measurement matrix $\bm{A}$.
For AS-ISTA, the learning rate for structural parameters $\beta_{r,k}^{(t)},\beta_{x,k}^{(t)}$ was $10^{-2}$ and
that for $\gamma^{(t)}$ was $10^{-4}$.
For AS-FISTA, the learning rate for structural parameters $\beta_{r,k}^{(t)},\beta_{x,k}^{(t)}$ was $5\times10^{-3}$,
that for $\beta_{z,k}^{(t)}$ was $10^{-2}$,
and that for $\gamma^{(t)}$ was $10^{-4}$.
The mini-batch size and the number of training iterations per layer for incremental learning were 100 and 50, respectively.
In the training process, we used STE \cite{ste}.
For the proposed methods, the initial values of structural parameters were $\beta_{r,1}^{(0)}=\beta_{x,2}^{(0)}=\beta_{z,1}^{(0)}=1$ and $\beta_{r,2}^{(0)}=\beta_{x,1}^{(0)}=\beta_{z,2}^{(0)}=-1$, 
and that for the learning rate was set similarly to ISTA.
The coefficient in $\widetilde{\|\bm{x}\|_1}$ and $\tilde{S}_\tau$ was $p=50$ for accurate approximation.
The meta learning rates for HGD-AS-ISTA were $\eta_r = \eta_x=10^{-1},\eta_\gamma=5\times10^{-9}$,
and those for HGD-AS-FISTA were $\eta_r = 10^{-1}, \eta_x=\eta_z=5\times10^{-2}, \eta_\gamma=5\times10^{-9}$.



\begin{figure}[tb]
  \centering
  \includegraphics[width=\linewidth]{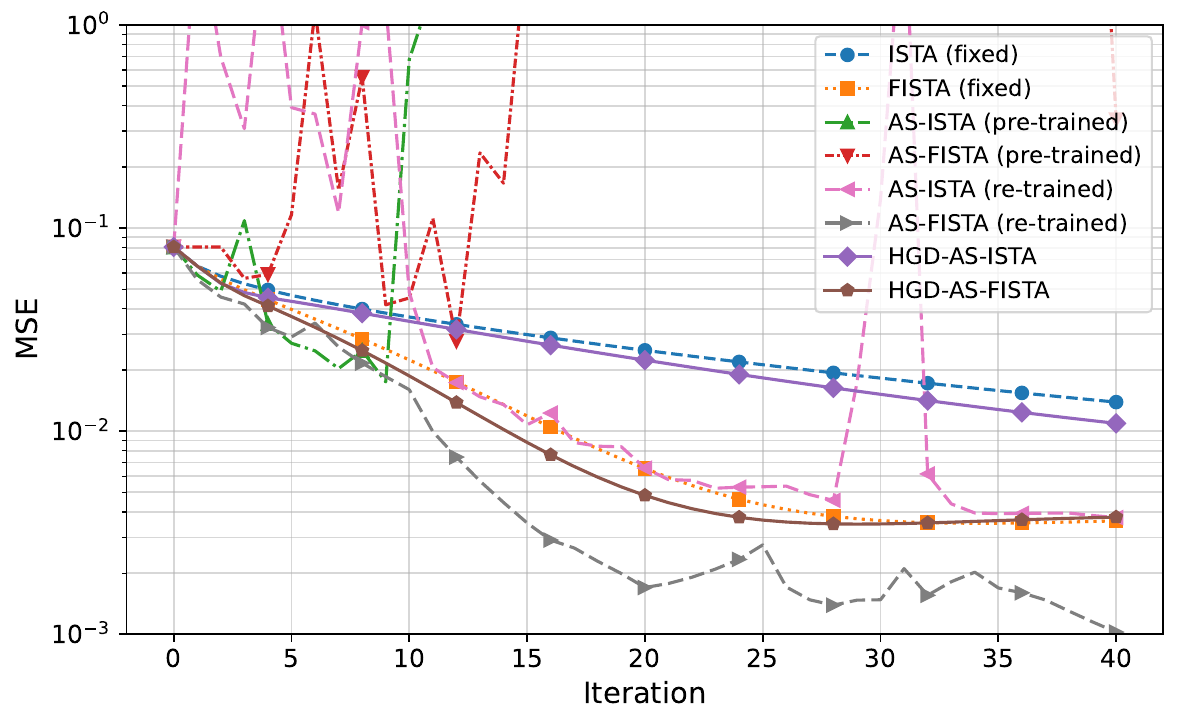}
  \caption{Comparison of MSE vs iteration.}
  \label{fig:mse}
\end{figure}
Figure \ref{fig:mse} shows the MSE values at each iteration.
AS-ISTA/FISTA performed poorly when using pre-trained parameters due to the change in the statistical properties of matrix $\bm{A}$, 
and required re-training to adapt to new environments.
On the other hand, 
the proposed HGD-AS-ISTA and HGD-AS-FISTA achieve better performance than the fixed cases 
while not requiring training data.

\begin{figure}[tb]
  \centering
  \subfigure[ISTA (fixed)]{%
    \includegraphics[width=0.30\linewidth]{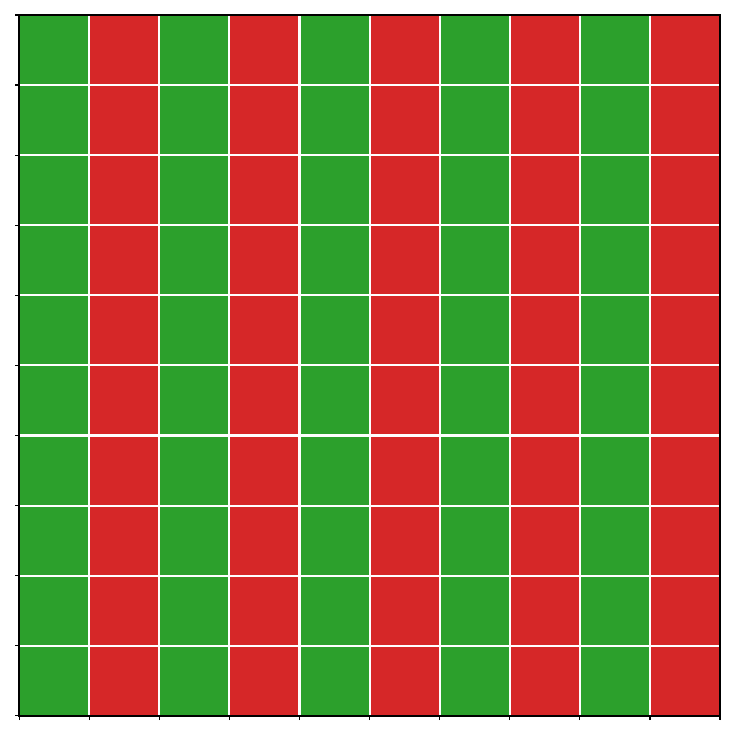}%
    \label{fig:heatmap_ista}
  }%
  \subfigure[AS-ISTA]{%
    \includegraphics[width=0.30\linewidth]{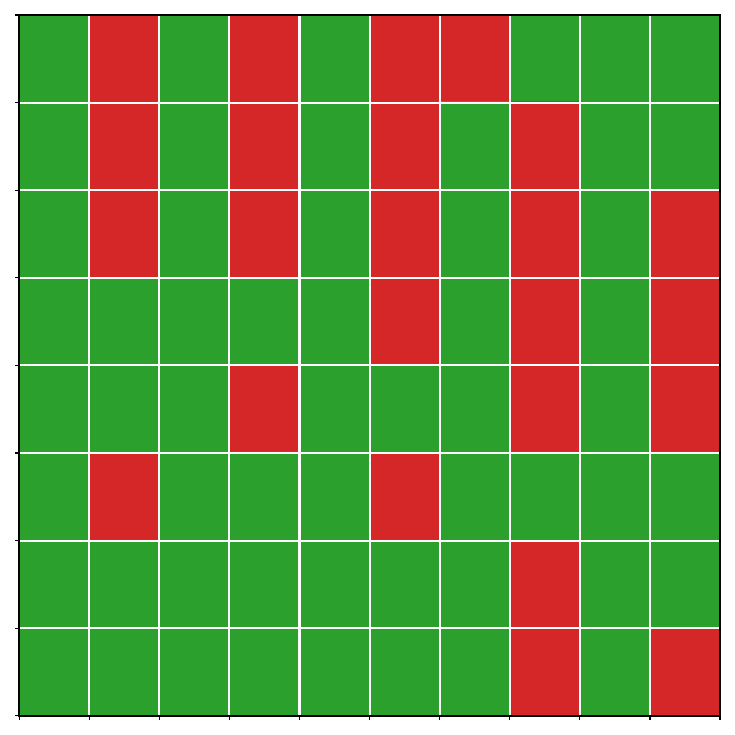}%
    \label{fig:heatmap_asista}
  }
  \subfigure[HGD-AS-ISTA]{%
    \includegraphics[width=0.30\linewidth]{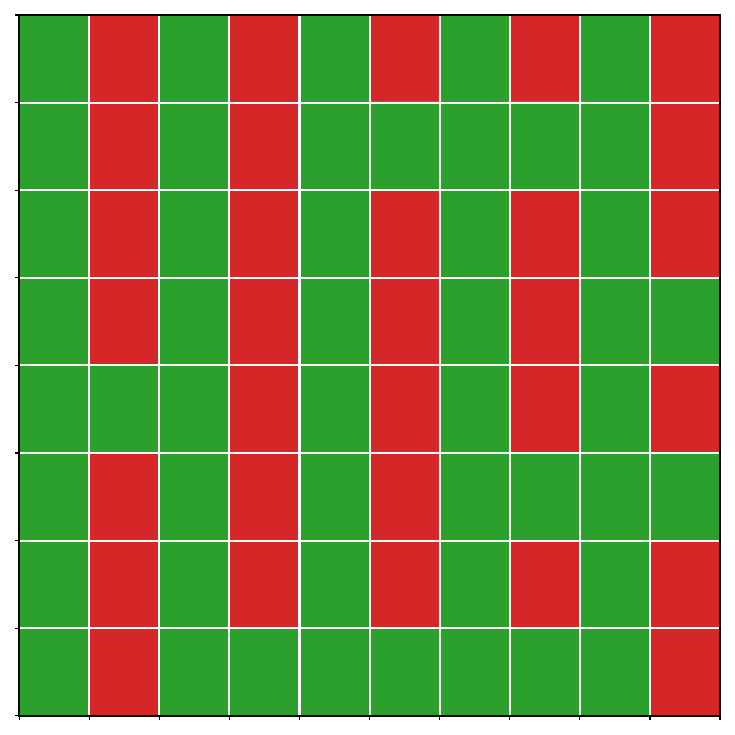}%
    \label{fig:heatmap_hgd_asista}
  }%
  \\
  \subfigure[FISTA (fixed)]{%
    \includegraphics[width=0.30\linewidth]{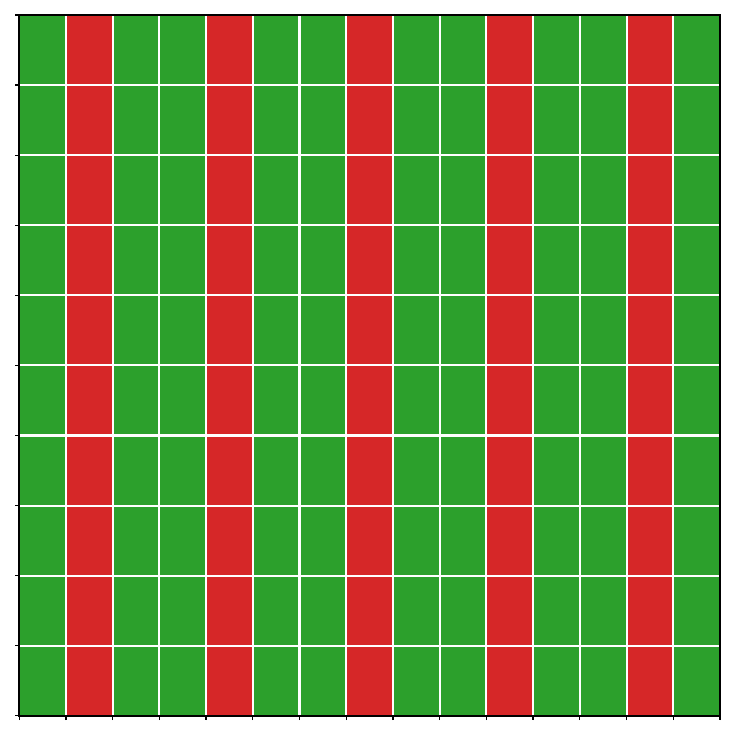}%
    \label{fig:heatmap_fista}
  }%
  \subfigure[AS-FISTA]{%
    \includegraphics[width=0.30\linewidth]{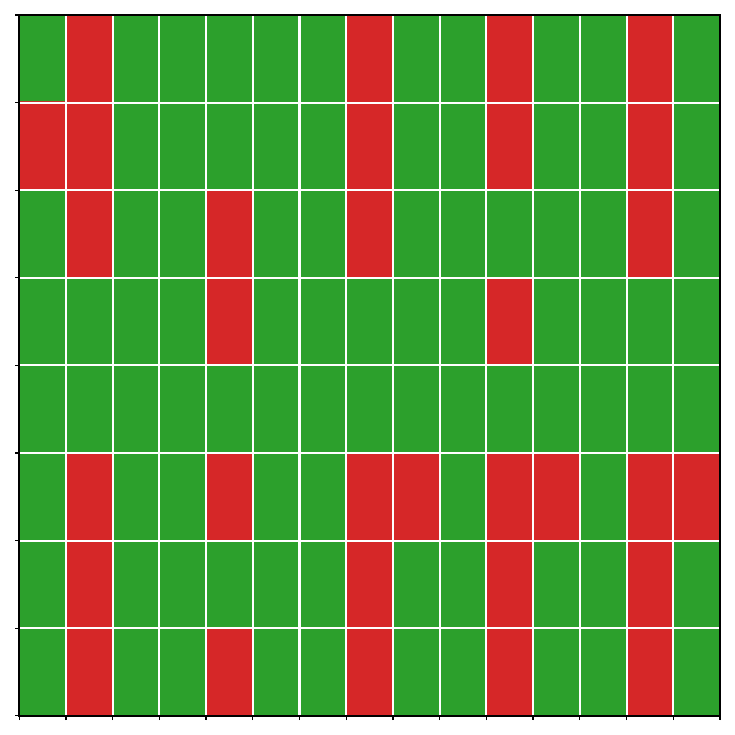}%
    \label{fig:heatmap_asfista}
  }%
  \subfigure[HGD-AS-FISTA]{%
    \includegraphics[width=0.30\linewidth]{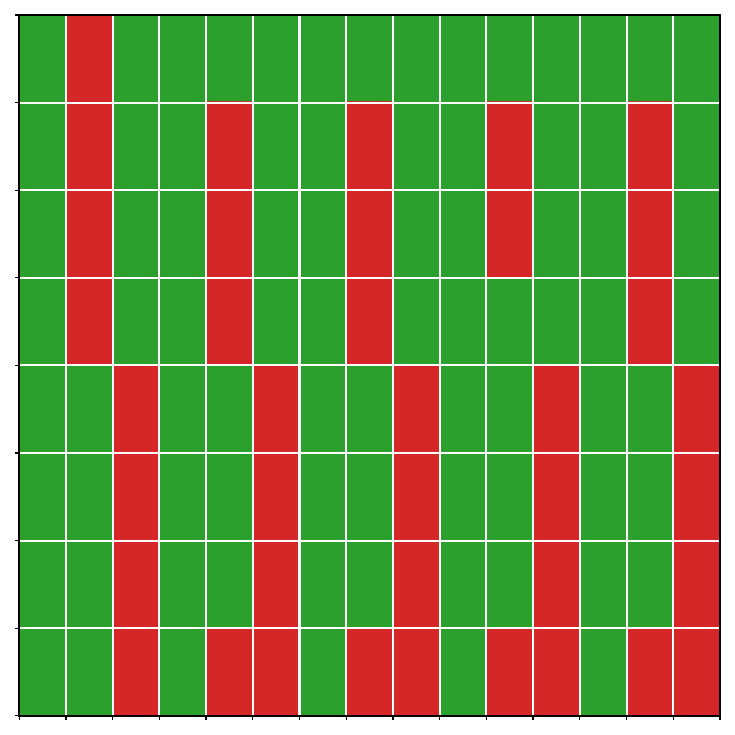}%
    \label{fig:heatmap_hgd_asfista}
  }
  \caption{Examples of architecture selected for each algorithm.}
  \label{fig:heatmaps}
\end{figure}

Figure \ref{fig:heatmaps} shows examples of the architecture selected for each algorithm.
Each block in the heatmaps represents the value of $w_{\cdot,1}^{(t)}$.
Green blocks are $w_{\cdot,1}^{(t)}=1$ and red blocks are $w_{\cdot,1}^{(t)}=0$.
For example, for ISTA, odd-numbered sequences represent the value of $w_{r,1}^{(t)}$ and even-numbered sequences represent the value of $w_{x,1}^{(t)}$.
The proposed method demonstrates appropriate structural choices, such as prioritizing gradient descent over shrinkage.

\begin{table}[t]
  \centering
  \caption{Execution time comparison.}
  \label{tab:execution_time}
  \begin{tabular}{|l|c|c|}
    \hline
    \textbf{Method} & \textbf{Training (ms)} & \textbf{Test per Signal (ms)} \\
    \hline
    ISTA (fixed) & --- & 2.44 \\
    \hline
    FISTA (fixed) & --- & 2.92 \\
    \hline
    AS-ISTA (re-trained) & 100129 & 8.90 \\
    \hline
    AS-FISTA (re-trained) & 124523 & 12.48 \\
    \hline
    HGD-AS-ISTA & --- & 20.61 \\
    \hline
    HGD-AS-FISTA & --- & 23.94 \\
    \hline
  \end{tabular}
\end{table}
Table \ref{tab:execution_time} shows the execution time of each method.
It includes the re-training time for AS-ISTA/FISTA, and the test time per signal.
The execution time of the proposed methods is larger than those of the fixed parameter methods 
due to the calculation of hypergradients for each signal.
However, considering that the re-training of AS-ISTA/FISTA takes over 100 seconds, 
the proposed methods avoid significant system downtime or estimation failure during environment transitions, 
making it more practical for real-time applications in non-stationary environments.

\section{Conclusion}
In this paper, we proposed HGD-AS-ISTA and HGD-AS-FISTA 
that introduce online parameter updates based on HGD into AS-ISTA/FISTA, 
compressed sensing algorithms that incorporate the idea of architecture search,
in order to eliminate the need for training data and time for deep unfolding.
Through computer experiments, 
we showed that the proposed methods have high adaptability to the execution environments.
Future work includes tracking performance evaluation when applying the proposed methods to practical wireless communication environments.

\appendices
\section{Hypergradient calculation for $\gamma^{(t)}$}
First, we show the hypergradient computation for $\gamma^{(t)}$ in HGD-AS-ISTA.
Specifically, we calculate
\begin{equation}
\frac{\partial \tilde{J}(\bm{x}^{(t+1)})}{\partial \gamma^{(t)}}
  = \frac{\partial \tilde{J}(\bm{x}^{(t+1)})}{\partial \bm{x}^{(t+1)}}
   \,\frac{\partial \bm{x}^{(t+1)}}{\partial \gamma^{(t)}} 
   =\nabla \tilde{J}(\bm{x}^{(t+1)}) ^\mathrm{T}
   \,\frac{\partial \bm{x}^{(t+1)}}{\partial \gamma^{(t)}}.
\label{eq:hypergradient_gammat}
\end{equation}

The former part $\nabla\tilde{J}(\bm{x}^{(t+1)})$ is
\begin{equation}
\begin{split}
\nabla\tilde{J}(\bm{x}^{(t+1)}) 
&= \frac{1}{2}\frac{d}{d\bm{x}}\left(\bm{y}^{\mathsf T}\bm{y} - \bm{y}^{\mathsf T}\bm{A}\bm{x} - \bm{x}^{\mathsf T}\bm{A}^{\mathsf T}\bm{y} + \bm{x}^{\mathsf T}\bm{A}^{\mathsf T}\bm{A}\bm{x}\right) \\
&= \bm{A}^\mathrm{T}(\bm{A}\bm{x}^{(t+1)}-\bm{y})+\lambda\left(2\sigma(p\bm{x}^{(t+1)})-\bm{1}\right).
\end{split}
\label{eq:nabla_tilde_J}
\end{equation}

For the latter part $\partial \bm{x}^{(t+1)}/\partial \gamma^{(t)}$ on the right-hand side of \eqref{eq:hypergradient_gammat},
substituting \eqref{eq:asista_x} gives
\begin{equation}
\begin{split}
\frac{\partial \bm{x}^{(t+1)}}{\partial \gamma^{(t)}} 
&= \frac{\partial}{\partial \gamma^{(t)}}\left(w_{x}^{(t)}f^{(t)}(\bm{r}^{(t)})+(1-w_{x}^{(t)})g^{(t)}(\bm{r}^{(t)})\right). \\
&=w_{x,1}^{(t)}\frac{\partial f^{(t)}(\bm{r}^{(t)})}{\partial\gamma^{(t)}}+w_{x,2}^{(t)}\frac{\partial g^{(t)}(\bm{r}^{(t)})}{\partial\gamma^{(t)}}.
\end{split}
\label{eq:partial_x_gamma}
\end{equation}
The functions $f^{(t)}, g^{(t)}$ and vector $\bm{r}^{(t)}$ include the parameter $\gamma^{(t)}$. 

The first term on the right-hand side of \eqref{eq:partial_x_gamma} can be computed as 
\begin{equation}
\frac{\partial f^{(t)}(\bm{r}^{(t)})}{\partial\gamma^{(t)}} = -\bm{A}^\mathrm{T}(\bm{A}\bm{r}^{(t)}-\bm{y})+\frac{\partial f^{(t)}(\bm{r}^{(t)})}{\partial\bm{r}^{(t)}}\frac{\partial \bm{r}^{(t)}}{\partial\gamma^{(t)}}.
\label{eq:partial_f_gamma}
\end{equation}
Each part can be calculated as follows:
\begin{equation}
\frac{\partial f^{(t)}(\bm{r}^{(t)})}{\partial\bm{r}^{(t)}} = \bm{I}-\gamma^{(t)}\bm{A}^\mathrm{T}\bm{A},
\label{eq:partial_f_r}
\end{equation}
\begin{align}
\frac{\partial \bm{r}^{(t)}}{\partial\gamma^{(t)}} &= -w_{r,1}^{(t)}\bm{A}^\mathrm{T}(\bm{A}\bm{x}^{(t)}-\bm{y})+w_{r,2}^{(t)}\lambda(\sigma(p(-\bm{x}^{(t)}-\gamma^{(t)}\lambda\bm{1})) \nonumber \\
& \quad -\sigma(p(\bm{x}^{(t)}-\gamma^{(t)}\lambda\bm{1}))).
\label{eq:partial_r_gamma}
\end{align}

For the second term on the right-hand side of \eqref{eq:partial_x_gamma}, 
the function $g^{(t)}(\bm{r}^{(t)})$ is not differentiable 
so we use the following differentiable approximation instead for subsequent calculations:
\[
S_{\tau}(q) \approx \frac{1}{p}\log(1+e^{p(q-\tau)}) - \frac{1}{p}\log(1+e^{p(-q-\tau)}) = \tilde{S}_{\tau}(q).
\]
This derivative can be expressed using the sigmoid function as 
$d\tilde{S}_{\tau}(q)/dq = \sigma(p(q-\tau)) + \sigma(p(-q-\tau))$.
By using the approximated function $\widetilde{g^{(t)}}(\bm{r}^{(t)}) = \tilde{S}_{\lambda\gamma^{(t)}}(\bm{r}^{(t)})$,
the derivative can be expressed as
\begin{align}
\frac{\partial \widetilde{g^{(t)}}(\bm{r}^{(t)})}{\partial\gamma^{(t)}} &\!= \!\lambda\!\left(\sigma(p(-\bm{r}^{(t)}-\gamma^{(t)}\lambda\bm{1}))\!-\!\sigma(p(\bm{r}^{(t)}-\gamma^{(t)}\lambda\bm{1}))\right) \nonumber \\
& \quad + \frac{\partial \widetilde{g^{(t)}}(\bm{r}^{(t)})}{\partial\bm{r}^{(t)}}\frac{\partial \bm{r}^{(t)}}{\partial\gamma^{(t)}},
\label{eq:partial_g_gamma}
\end{align}
The term $\partial \bm{r}^{(t)}/\partial\gamma^{(t)}$ is as in \eqref{eq:partial_r_gamma} and 
another part can be calculated as follows:
\begin{equation}
\frac{\partial \widetilde{g^{(t)}}(\bm{r}^{(t)})}{\partial\bm{r}^{(t)}} \!= \!\mathrm{diag}\!\left[\sigma(p(\bm{r}^{(t)}\!-\!\gamma^{(t)}\lambda\bm{1}))\!+\!\sigma(p(-\bm{r}^{(t)}\!-\!\gamma^{(t)}\lambda\bm{1}))\!\right].
\label{eq:partial_g_r}
\end{equation}

By substituting \eqref{eq:nabla_tilde_J}--\eqref{eq:partial_g_r} into \eqref{eq:hypergradient_gammat},
the hypergradient for $\gamma^{(t)}$ can be obtained.

\section{Hypergradient calculation for $w_r^{(t)}$ and $w_x^{(t)}$}
We show the computation of the hypergradient for $\beta_{r,k}^{(t)}$,
\begin{equation}
\frac{\partial\tilde{J}(\bm{x}^{(t+1)})}{\partial \beta_{r,k}^{(t)}} 
=\nabla \tilde{J}(\bm{x}^{(t+1)}) ^\mathrm{T} \frac{\partial \bm{x}^{(t+1)}}{\partial \beta_{r,k}^{(t)}}.
\label{eq:wr}
\end{equation}
For any $w_{\cdot,k}^{(t)}$ that can be expressed as \eqref{eq:expr},
\begin{equation}
  \frac{\partial w_{\cdot, 1}^{(t)}}{\partial \beta_{\cdot,1}^{(t)}}=w_{\cdot,1}^{(t)}w_{\cdot,2}^{(t)} = \frac{\partial w_{\cdot,2}^{(t)}}{\partial \beta_{\cdot,2}^{(t)}},
  \frac{\partial w_{\cdot,2}^{(t)}}{\partial \beta_{\cdot,1}^{(t)}}=-w_{\cdot,1}^{(t)}w_{\cdot,2}^{(t)} = \frac{\partial w_{\cdot,1}^{(t)}}{\partial \beta_{\cdot,2}^{(t)}}.
  \label{eq:wtobeta}
\end{equation}

Consider the case of $k=1$.
For the latter part $\partial \bm{x}^{(t+1)}/\partial \beta_{r,1}^{(t)}$,
substituting the update equation \eqref{eq:asista_x} gives
\begin{equation}
  \frac{\partial \bm{x}^{(t+1)}}{\partial \beta_{r,1}^{(t)}}=w_{x,1}^{(t)}\frac{\partial f^{(t)}(\bm{r}^{(t)})}{\partial \bm{r}^{(t)}}\frac{\partial \bm{r}^{(t)}}{\partial \beta_{r,1}^{(t)}}
  +w_{x,2}^{(t)}\frac{\partial \widetilde{g^{(t)}}(\bm{r}^{(t)})}{\partial \bm{r}^{(t)}}\frac{\partial \bm{r}^{(t)}}{\partial \beta_{r,1}^{(t)}},
  \label{eq:partial_x_betar}
\end{equation}
Regarding the uncalculated term $\partial \bm{r}^{(t)}/\partial \beta_{r,1}^{(t)}$ on the right-hand side of \eqref{eq:partial_x_betar}, by using \eqref{eq:wtobeta}, 
\begin{align}
  \frac{\partial \bm{r}^{(t)}}{\partial \beta_{r,1}^{(t)}}
  &=\frac{\partial w_{r,1}^{(t)}}{\partial \beta_{r,1}^{(t)}}f^{(t)}(\bm{x}^{(t)})+\frac{\partial w_{r,2}^{(t)}}{\partial \beta_{r,1}^{(t)}}\widetilde{g^{(t)}}(\bm{x}^{(t)}) \nonumber \\
  &= w_{r,1}^{(t)}w_{r,2}^{(t)}(f^{(t)}(\bm{x}^{(t)})-\widetilde{g^{(t)}}(\bm{x}^{(t)})).
  \label{eq:partial_r_betar}
\end{align}

By substituting \eqref{eq:nabla_tilde_J}, \eqref{eq:partial_f_r}, \eqref{eq:partial_g_r}, \eqref{eq:partial_x_betar}, and \eqref{eq:partial_r_betar} into \eqref{eq:wr},
the hypergradient for $\beta_{r,1}^{(t)}$ can be calculated.
Furthermore, $\partial\tilde{J}(\bm{x}^{(t+1)})/\partial \beta_{r,2}^{(t)} = -\partial\tilde{J}(\bm{x}^{(t+1)})/\partial \beta_{r,1}^{(t)}$ 
since \eqref{eq:wtobeta} holds.

The hypergradient for $\beta_{x,k}^{(t)}$, 
\begin{equation}
  \frac{\partial \tilde{J}(\bm{x}^{(t+1)})}{\partial \beta_{x,k}^{(t)}}=\frac{\partial \tilde{J}(\bm{x}^{(t+1)})}{\partial \bm{x}^{(t+1)}}\frac{\partial \bm{x}^{(t+1)}}{\partial \beta_{x,k}^{(t)}}, \ (k=1,2),
\end{equation}
can be directly calculated.
For $k=1$, 
\begin{align}
  \frac{\partial \bm{x}^{(t+1)}}{\partial \beta_{x,1}^{(t)}} 
  &= \frac{\partial w_{x,1}^{(t)}}{\partial \beta_{x,1}^{(t)}}f^{(t)}(\bm{r}^{(t)})+\frac{\partial w_{x,2}^{(t)}}{\partial \beta_{x,1}^{(t)}}\widetilde{g^{(t)}}(\bm{r}^{(t)}) \nonumber \\ 
  &= w_{x,1}^{(t)}w_{x,2}^{(t)} (f^{(t)}(\bm{r}^{(t)})-\widetilde{g^{(t)}}(\bm{r}^{(t)})), 
\end{align}
and the resulting hypergradients can be derived by substituting similarly to the case of $\beta_{r,k}^{(t)}$. 
Also, $\partial\tilde{J}(\bm{x}^{(t+1)})/\partial \beta_{x,2}^{(t)} = -\partial\tilde{J}(\bm{x}^{(t+1)})/\partial \beta_{x,1}^{(t)}$ holds for $k=2$.

\section{Hypergradient calculation for HGD-AS-FISTA}
In the case of HGD-AS-FISTA, 
the derivation of hypergradients for $\gamma^{(t)},\beta_{r,k}^{(t)},\beta_{x,k}^{(t)}$ is largely similar to that of the HGD-AS-ISTA, 
but, since it involves $\bm{z}^{(t+1)}$, the following factor is required to be applied:
\begin{equation}
  \frac{\partial \bm{z}^{(t+1)}}{\partial \cdot^{(t)}} = \left(w_{z,1}^{(t)}\frac{s^{(t+1)}+s^{(t)}-1}{s^{(t+1)}}+w_{z,2}^{(t)}\right)\frac{\partial \bm{x}^{(t+1)}}{\partial\cdot^{(t)}}.
\end{equation}

The hypergradient for $\beta_{z,k}^{(t)}$, i.e., 
\begin{equation}
  \frac{\partial \tilde{J}(\bm{z}^{(t+1)})}{\partial \beta_{z,k}^{(t)}}=\frac{\partial \tilde{J}(\bm{z}^{(t+1)})}{\partial \bm{z}^{(t+1)}}\frac{\partial \bm{z}^{(t+1)}}{\partial \beta_{z,k}^{(t)}} \ (k=1,2)
\end{equation}
can be directly calculated: 
\begin{align}
  \frac{\partial \bm{z}^{(t+1)}}{\partial \beta_{z,1}^{(t)}} 
  &= \frac{\partial w_{z,1}^{(t)}}{\partial \beta_{z,1}^{(t)}}h(\bm{x}^{(t+1)},\bm{x}^{(t)},s^{(t+1)},s^{(t)})+\frac{\partial w_{z,2}^{(t)}}{\partial \beta_{z,1}^{(t)}}\bm{x}^{(t+1)} \nonumber\\ 
  &= w_{z,1}^{(t)}w_{z,2}^{(t)}(h(\bm{x}^{(t+1)},\bm{x}^{(t)},s^{(t+1)},s^{(t)})-\bm{x}^{(t+1)}) \nonumber\\ 
  &= w_{z,1}^{(t)}w_{z,2}^{(t)}\frac{s^{(t)}-1}{s^{(t+1)}}(\bm{x}^{(t+1)}-\bm{x}^{(t)}),
\end{align}
and $\partial\tilde{J}(\bm{z}^{(t+1)})/\partial \beta_{z,2}^{(t)} = -\partial\tilde{J}(\bm{z}^{(t+1)})/\partial \beta_{z,1}^{(t)}$ holds.

\end{document}